\documentclass[aip,reprint]{revtex4-1}

\usepackage[utf8]{inputenc}
\usepackage{graphicx}
\usepackage{epstopdf}
\usepackage{siunitx}
\usepackage{mathtools}
\usepackage{color}

\begin{document}

 \title{Dipole-dipole correlations and the Debye process in the dielectric response of non-associating glass forming liquids}
 
 \author{Florian Pabst}
\author{Andreas Helbling}
\noaffiliation
\author{Jan Gabriel}
\noaffiliation
\noaffiliation
\author{Peter Weigl}
\noaffiliation
\author{Thomas Blochowicz}
\noaffiliation
\affiliation{TU Darmstadt, Institute of Condensed Matter Physics, 64289 Darmstadt, Germany}
 
\date{\today}

\begin{abstract}
The non-exponential shape of the $\alpha$-process observed in supercooled liquids is considered as one of the hallmarks of glassy dynamics and has thus been under study for decades, but is still poorly understood. For a polar van der Waals liquid, we show here -- in line with a recent theory -- that dipole-dipole correlations give rise to an additional process in the dielectric spectrum slightly slower than the $\alpha$-relaxation, which renders the resulting combined peak narrower than observed by other experimental techniques. This is reminiscent of the Debye process found in monohydroxy alcohols. The additional peak can be suppressed by weakening the dipole-dipole interaction via dilution with a nonpolar solvent.    
\end{abstract}
 
\maketitle 

The hallmarks of glassy dynamics are sometimes characterized by the so-called ``three nons'',\cite{dyre2006colloquium} which refer to the non-Arrhenius temperature dependence of characteristic time constants, the non-exponential time dependence of the structural relaxation and the non-linear behavior in response to small temperature changes.\cite{giordano1996non, wang2008calorimetric, angell2000relaxation,cavagna2009supercooled,donth2013glass} Despite decades of research and various proposed theories, no consensus about the origin of these features has been reached up to now. Concerning the spectral shape of the $\alpha$-relaxation, which is the focus of the present paper, previous studies have tried to point out scaling relations \cite{dixon1990scaling, dendzik1997universal, paluch1998parameters} or otherwise common features of the line shape.\cite{blochowicz2003susceptibility} In some works a connection of secondary relaxations with the spectral shape of the $\alpha$-process is highlighted,\cite{ngai2011relaxation} while others point out that the line shape may be universal altogether \cite{nielsen2009prevalence} or at least similar for a subclass of particularly ``simple'' liquids. \cite{niss2018perspective}   


In any case, the structural relaxation is found to be non-exponential, which is most often described by the empirical Kohlrausch-Williams-Watts (KWW) equation in the time domain:\cite{williams1970non} $f(t) = e^{-\left(\tau/t\right)^{\beta_K}}$, with the correlation time $\tau$ and a stretching parameter $0<\beta_{K}<1$.
Although initially it was believed that a single-exponential relaxation ($\beta_{K} =1$) or, equivalently, a Debye peak $\left(\epsilon''(\omega) = \Delta \epsilon/(1+(\i\omega\tau)\right)$ in the dielectric loss reflects the structural relaxation in certain alcohols,\cite{debye1929polar} it later became clear that the structural $\alpha$-relaxation is non-exponential in all supercooled liquids and that monohydroxy alcohols are just exceptional in the sense that an \emph{additional} Debye-like process is visible in dielectric measurements usually seen at longer times or lower frequencies.\cite{bohmer2014structure}  The occurrence of such an additional Debye process is usually ascribed to the relaxation of an average end-to-end dipole moment of transient chains formed by hydrogen bonding in monohydroxy alcohols.\cite{gainaru2010nuclear}

However, based on the work of Dean, \cite{dean1996langevin} Kawasaki\cite{kawasaki1994stochastic} and Cugliandolo {\it et al.} \cite{cugliandolo2015stochastic}, D\'ejardin {\it et al.}\cite{dejardin2019linear} recently showed that an additional process should arise in the dielectric loss spectrum, whenever the Kirkwood correlation factor $g_K$, which describes static correlations between interacting dipoles,\cite{boettcher1978theory1} exceeds unity, \emph{without} explicitly referring to the existence of H-bonded structures. At first glance this may be surprising, as, e.g., in polar van der Waals liquids no additional slow relaxation was reported so far and the notion that \emph{dynamic} cross-correlations 
may in general be negligible is fairly wide spread.\cite{boettcher1978theory1, williams1972molecular, kivelson1975theory} 
Remarkably, however, a correlation between stretching parameter and relaxation strength was reported recently for a set of 88 glass formers,\cite{paluch2016universal} indicating a more narrow relaxation peak in highly dipolar liquids. Thus, in principle an additional collective and possibly narrow or Debye-like process could be present in polar liquids, albeit its strong overlap with the structural $\alpha$-relaxation might prevent a clear distinction of both processes. In fact, quite the same holds true in several monohydroxy alcohols, where previously a combination of depolarized dynamic light scattering (DDLS) measurements with broadband dielectric data allowed to disentangle both processes.
 It was demonstrated that indeed BDS and DDLS spectra  are identical except for the Debye contribution in many alcohols\cite{boehmer2019influence,gabriel2017debye,gabriel2018nature}.

Therefore, in the following we address the question, if such an additional slow collective process can indeed be distinguished in a polar van der Waals liquid, namely tributyl phosphate (TBP), where no H-bonds are present and which exhibits $g_{K} > 1$ at low temperatures.\cite{kirkwoodTBP} An indication that this could be the case is already given by the reported discrepancy between the non-exponential parameter $\beta_{\text{K}}$ obtained by BDS and by calorimetry.\cite{wu2017presence} To achieve this goal, BDS and DDLS data are combined in a broad frequency and temperature range and the same method of analysis is applied as previously used for monohydroxy alcohols.\cite{gabriel2017debye,gabriel2018nature,gabriel2018depolarized} In addition, dilution experiments are performed on the polar liquid and it is demonstrated that intermolecular dipole correlations can indeed be suppressed. 

Tributyl phosphate (Aldrich, $>$99\%) was filtered into the light scattering sample cell by using 200\,nm syringe filters, the dielectric samples were prepared without further purification. Depolarized dynamic light scattering was performed with a photon correlation setup described elsewhere in detail.\cite{gabriel2018depolarized} The obtained intensity autocorrelation function was transformed into the electric field correlation function via the Siegert relation for partially heterodyne signals\cite{pabst2017molecular} and afterwards Fourier transformed using the Filon integration rule to allow for a direct comparison between the generalized light scattering susceptibility and the dielectric loss.\cite{gabriel2018depolarized}    BDS measurements were performed with a Novocontrol Alpha-N Analyzer in combination with a home-built time domain setup.\cite{gabriel2018depolarized}
Special care was taken to calibrate temperature in the different setups in order to achieve an overall accuracy of $\pm$0.5~K.

\begin{figure}[h]
   \includegraphics[width=0.45\textwidth]{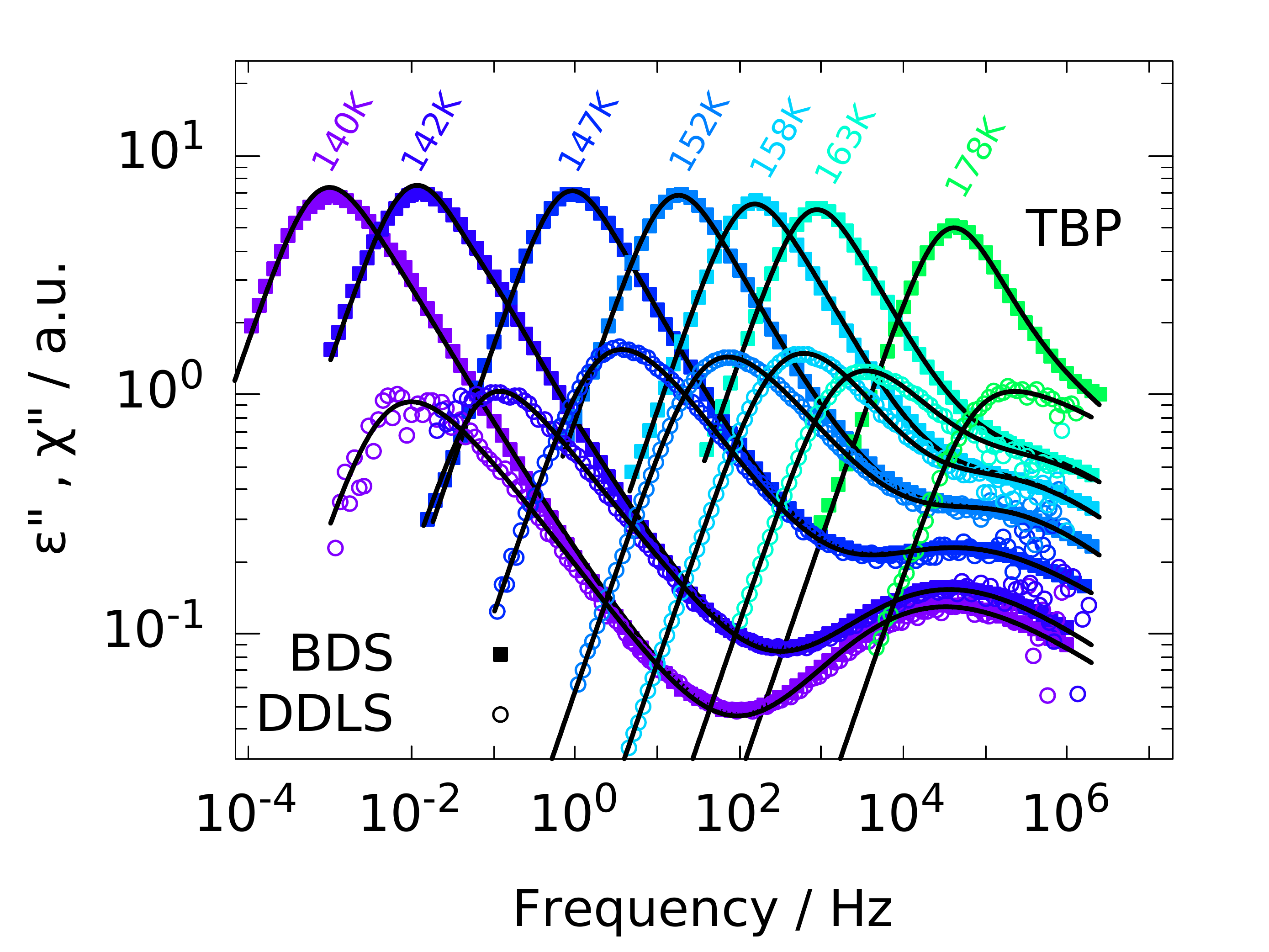}
   \caption{BDS and DDLS data of TBP for selected temperatures in the supercooled regime. DDLS data are scaled vertically to overlap with the BDS data at high frequencies. Large deviations are seen for the shape and position of the main process between BDS and DDLS. Black solid lines are fits, see text for details.} 
  \label{fig_BDS+PCS}
\end{figure}
Figure \ref{fig_BDS+PCS} shows BDS and DDLS loss spectra of neat TBP in susceptibility representation. Since the absolute height of the DDLS data is not experimentally accessible in a straight forward manner, the DDLS data are shifted vertically so that they overlap with the BDS data in the high frequency region beyond the $\alpha$-relaxation peak at each respective temperature.

\begin{figure}[h]
   \includegraphics[width=0.45\textwidth]{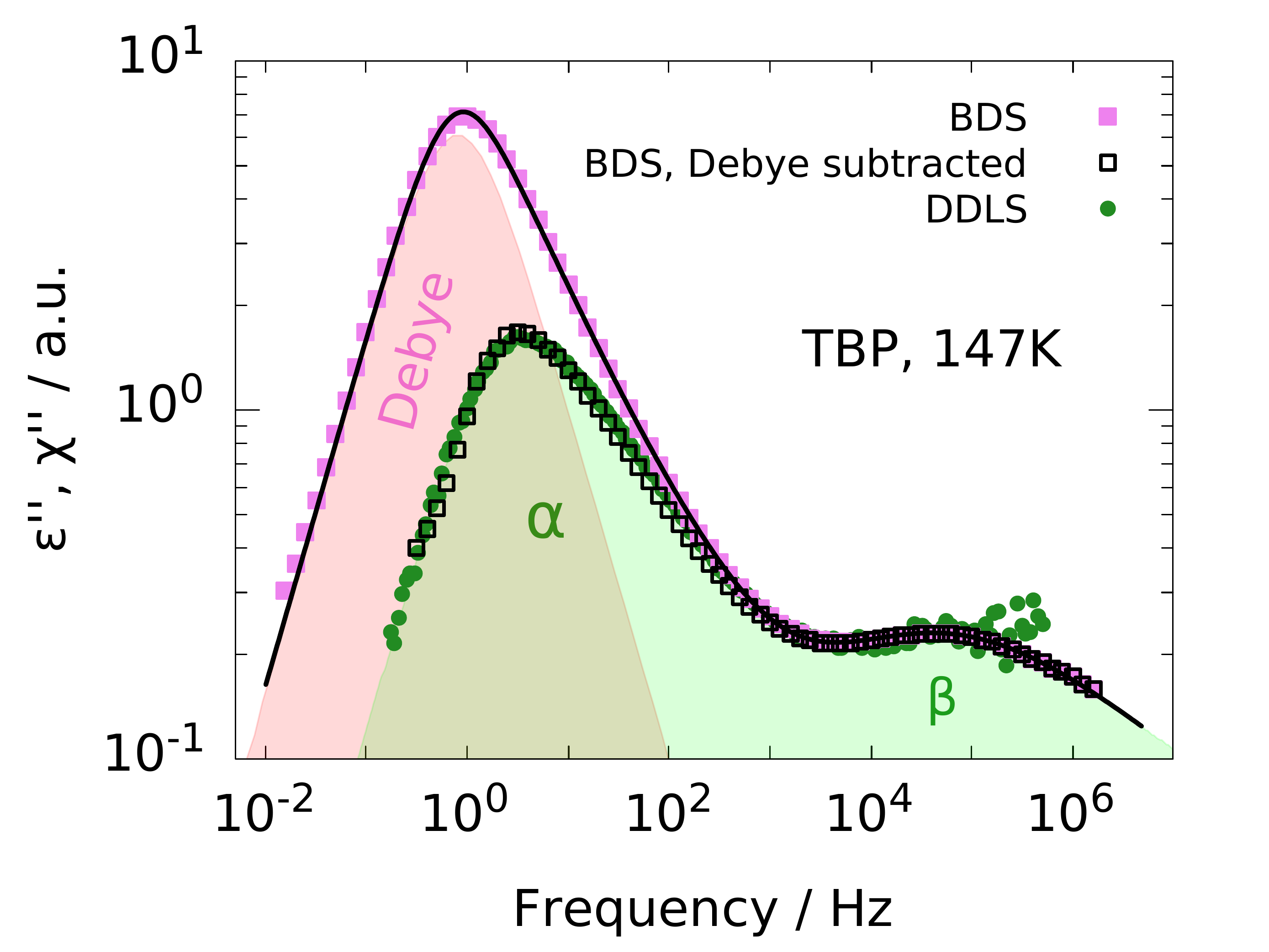}
   \caption{DDLS and BDS data of neat TBP at 147~K. In addition to the $\alpha$- and $\beta$-process present in the DDLS data, a Debye process is visible in the BDS spectra. BDS data are also shown with this Debye process subtracted to visualize the agreement of the remaining spectra with the DDLS data.} 
  \label{fig_fit}
\end{figure}
It can be seen that in the high frequency region, where a pronounced secondary relaxation is visible, both data sets have exactly the same shape. Deviations occur at low frequencies in the region of the main peak, which is located at lower frequencies and is more intense in BDS than in DDLS. 
As mentioned before this phenomenon is reminiscent of the situation found in monohydroxy alcohols, where a slow Debye process is found in BDS in addition to the $\alpha$-relaxation, the latter being identical in shape and position in both methods, whereas the Debye process is either not\cite{gabriel2017debye} or only very weakly\cite{gabriel2018nature} visible in DDLS and rheology.\cite{gainaru2014shear} 
In order to quantitatively check, if such a picture of an additional Debye process in the BDS spectra is in accordance with the TBP data, first, the DDLS data are fitted with a model containing an $\alpha$- and a $\beta$-process, which is described elsewhere in detail.\cite{gabriel2018nature} Then -- in order to describe the BDS data at the same temperature -- only a Debye process is added to the resulting DDLS fit function. As can be seen in Fig.~\ref{fig_fit}, this approach is able to describe the BDS data perfectly. 

In this figure the BDS data are also shown with the slow Debye process subtracted to visualize the agreement of the dielectric $\alpha$- and $\beta$-relaxation with the DDLS data. This fitting procedure works at all investigated temperatures, as can be seen in Fig.~\ref{fig_BDS+PCS}.  The implication is that on top of the $\alpha$-process, which is present in BDS and DDLS data alike, an intense Debye process is visible only in BDS, like it is the case in monohydroxy alcohols. As mentioned before, the occurrence of the Debye process in monohydroxy alcohols is usually ascribed to the formation of H-bonded supramolecular structures. However, since no H-bonds are present in TBP, this reasoning is not applicable here. 

The theory of Dejardin {\it et al.},\cite{dejardin2019linear} however, predicts an additional process in the dynamic susceptibility of dipolar assemblies in the more general case, whenever the Kirkwood correlation factor $g_K$ exceeds unity, that means when the dipoles tend to align parallel. In cases where $g_K<1$, i.e., with a preferentially antiparallel alignment of the dipoles, no additional process appears. Although the  approximations made while deriving the D\'ejardin formula prevent a quantitative comparison of theory and experiment, qualitatively, an additional slow process in the dielectric spectra would be expected also for TBP, since $g_K > 1$ at low temperatures.\cite{kirkwoodTBP} Therefore we tentatively attribute the Debye process only visible in the BDS data to arising from intermolecular dipolar interactions in TBP and we will further test this notion in the following. 

\begin{figure}[h]
   \includegraphics[width=0.45\textwidth]{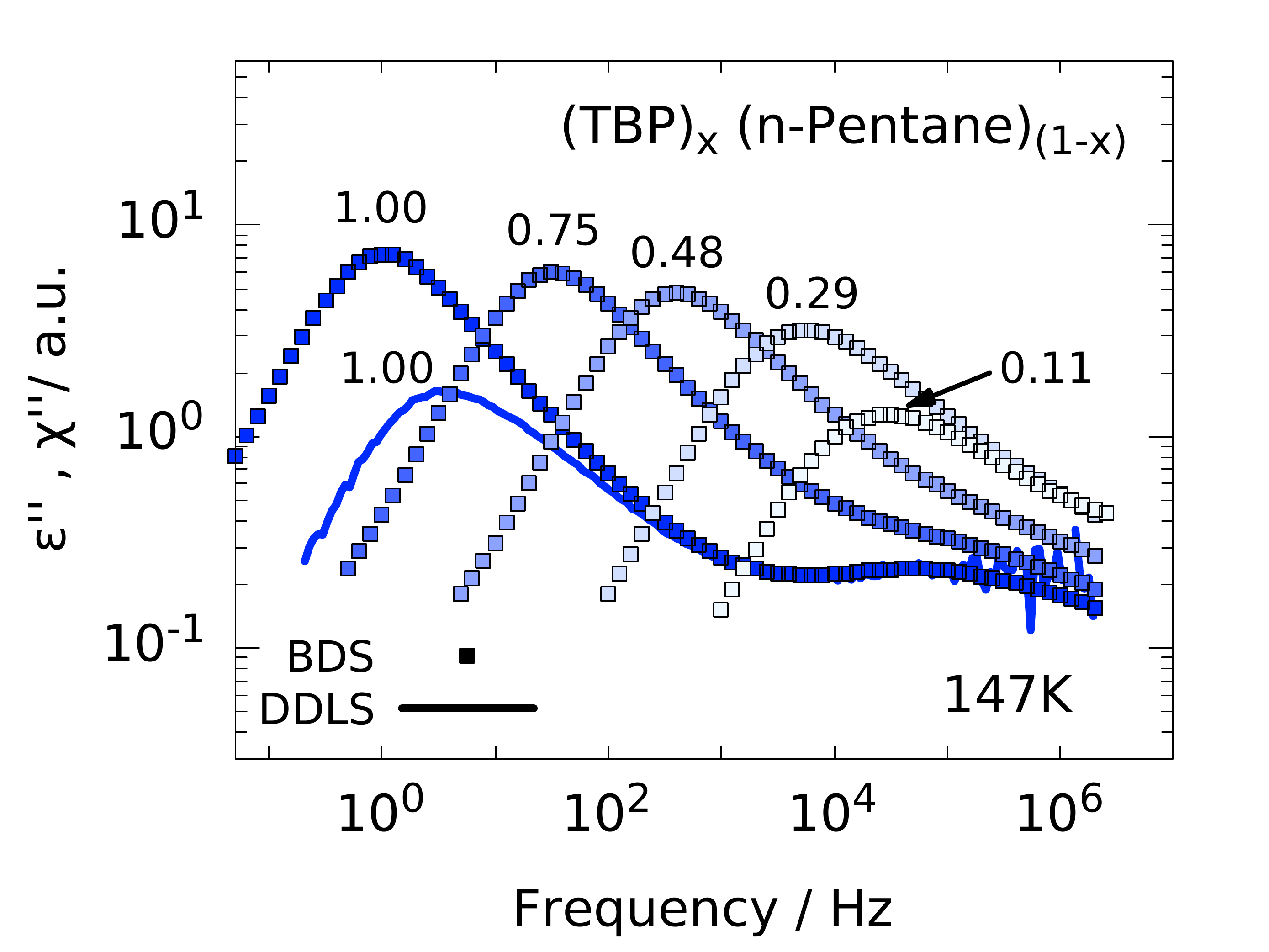}
   \caption{BDS data of TBP/n-pentane mixtures with mole fractions as indicated, measured at 147~K. DDLS data for neat TBP are shown for comparison as solid line. Besides the acceleration of the dynamics with decreasing mole fraction of TBP, a broadening of the main process is clearly visible.} 
  \label{fig_mix}
\end{figure}

The strength of the dipolar interactions should decrease when TBP is diluted with a non-polar solvent through a separation of the TBP molecules, thus leading to 
a vanishing of the additional Debye process upon dilution. If this is true, then, below a certain low concentration of TBP in the non-polar solvent, the shape of the main peak in the BDS data of the mixture should resemble the shape of the DDLS data. To test this hypothesis, we chose n-pentane (Arcos, 99+\%) as the non-polar solvent, because of its full miscibility with TBP over the whole concentration range, its low melting point of approximately 140~K and its approximated glass transition temperature $T_{g}$ of around 105~K.\cite{Tg_idealMix} The latter is important, because it is known that in binary mixtures with too large a difference in $T_{g}$ of the components, concentration fluctuations may increase the width of the $\alpha$-process of the mixture upon addition of one component, while it decreases again on further increasing the concentration beyond an equimolar composition of the components.\cite{shears1973molecular} Because the difference in $T_{g}$ of $\approx$35~K in the TBP/n-pentane mixture is quite small and we did not observe such a concentration dependent shape variation, we rule out concentration fluctuations to be the reason for broadening of the main process of the mixture. Instead, the results of the BDS measurements on the mixtures with molar ratios down to 11~mol~\% TBP are shown for the temperature of 147~K in figure \ref{fig_mix}. We note that no loss peak could be detected for neat n-pentane down to the instrumental resolution. 

\begin{figure}[h]
   \includegraphics[width=0.45\textwidth]{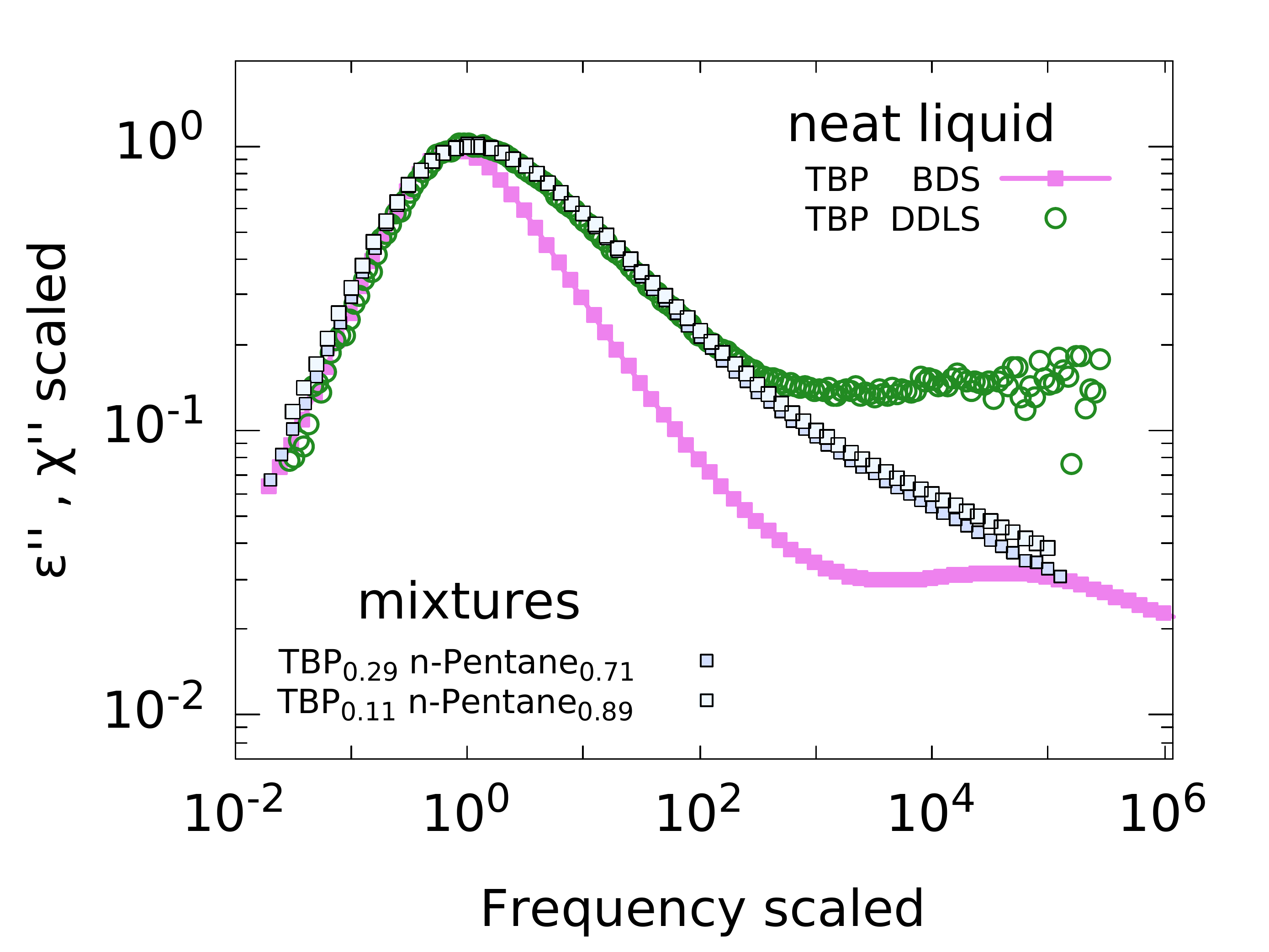}
   \caption{Data of neat (BDS+DDLS) and mixed TBP (BDS), scaled to overlap in peak height and on the low frequency flank. The $\alpha$-process of the two low-TBP-concentration mixtures measured by BDS and neat TBP measured by DDLS have the same spectral shape. By contrast, neat TBP measured by BDS exhibits a markedly narrower main process.} 
  \label{fig_master}
\end{figure}

It can be seen in this figure that with increasing concentration of n-pentane the main peak shifts to higher frequencies and decreases in intensity. Both effects are, however, expected since on the one hand n-pentane has a lower glass transition temperature than TBP and acts thus as a ``plasticizer'' in the mixture, accelerating the TBP dynamics, and on the other hand the mixture is getting less polar with increasing n-pentane concentration, thus decreasing the dielectric strength. However, an other effect is clearly visible by eye from Fig.~\ref{fig_mix}: As the concentration of TBP decreases, the width of the main peak increases. Provided this broadening is due to the vanishing of the additional interaction induced Debye process, there should be a concentration of TBP at which this process is no longer detectable and thus for even lower TBP concentrations, no further broadening should occur. Furthermore, the remaining $\alpha$-process should have the same spectral shape as detected by DDLS, since the DDLS spectrum resembles the BDS spectrum without the Debye peak, as shown in Fig.~\ref{fig_fit}. This notion is tested in Fig.~\ref{fig_master}, where the data of neat TBP measured by BDS and DDLS and mixtures of TBP and n-pentane with 29 and 11 mol-\% TBP are shown, scaled in such way that the peak heights and the low frequency flank of each data set coincides. Temperatures are chosen in such way that the main peak is located at around 1~Hz at this respective temperatures.      

It can be seen from Fig.~\ref{fig_master} that for the TBP/n-pentane mixtures with 29 and 11 mol~\% TBP the spectral shapes are the same, indicating indeed a limiting concentration of TBP, below which no further broadening occurs. In addition, the peak widths resemble the one of the DDLS data of neat TBP, whereat the BDS spectrum of neat TBP is markedly narrower.
We note, that it was already observed by Wu {\it et al.}\cite{wu2017presence} that the spectral width of BDS data of TBP is markedly narrower than the one obtained by differential scanning calorimetry (DSC). More precisely, they found the spectral width of the BDS peak in terms of the $\beta_{\text{K}}$ parameter to be 0.73 and for the DSC spectrum $\beta_{\text{K}}$ = 0.48.
To compare the spectral shape of our DDLS data or the low concentration TBP mixtures, respectively, with the one from DSC measurements, we used the same procedure as done by Wu {\it et al.} for comparing BDS with DSC: We fit our data with Havriliak-Negami equation (HN):\cite{havriliak1967complex}
\begin{equation}
\epsilon^*(\omega) = \frac{\Delta \epsilon}{(1+(\i\omega\tau)^a)^b}+\epsilon_{\infty}
\label{eq_HN}
\end{equation}
and transform the HN paramters a,b into $\beta_{\text{K}}$ via $\beta_{\text{K}} = (a\cdot b)^{0.813}$. In doing so we obtain a $\beta_{\text{K}}$ of 0.49, which is in very good agreement with 0.48 found by DSC. All in all, these findings indicate that in the BDS spectrum of neat TBP an additional Debye process is present, which can be suppressed by diluting TBP in order to minimize the interaction between TBP molecules. The resulting spectral shape in the dilute regime is identical with the spectral shape found for neat TBP with DDLS and DSC, thus exhibiting the real shape of the $\alpha$-process. 

These findings can be put in a more general context, if one considers the recent finding by Paluch {\it et al.}\cite{paluch2016universal} that for a large number of van der Waals molecular glass formers the stretching parameter $\beta_{\text{K}}$ of the $\alpha$-relaxation is correlated with the dielectric strength $\Delta \epsilon$ near the glass transition temperature $T_g$. This means that the stronger the dielectric loss of the molecule is, the narrower is its dielectric $\alpha$-peak. The authors explained this finding with the dipole-dipole interaction contribution to the attractive intermolecular potential, which enhances its harmonicity and thus making the $\alpha$-loss peak narrower and the $\beta_{\text{K}}$  parameter larger, respectively. How different experimental techniques reflect these differences in intermolecular potential, however, remains unclear.

\begin{figure}[h]
   \includegraphics[width=0.45\textwidth]{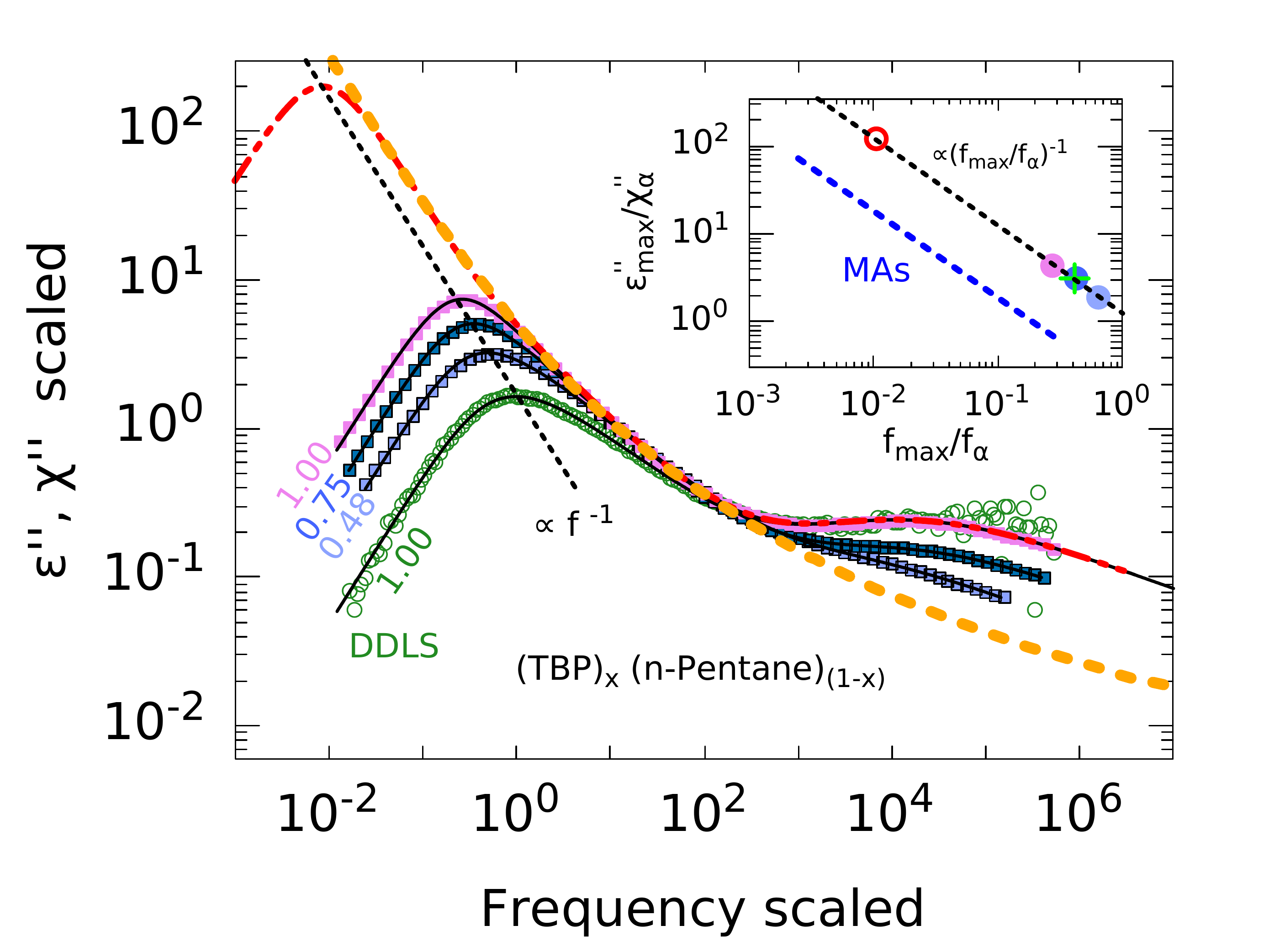}
   \caption{BDS data of TBP/n-pentane mixtures with mole fractions as indicated and DDLS data of neat TBP. All data shifted to superimpose on the envelope\cite{gainaru2019spectral} (orange dashed line). Solid black lines are fits and the red dashed line represents a hypthetical Debye process, see text for details. Inset: The peak height ratio $\epsilon''_{\text{max}}/\chi''_{\text{max}}$ as a function of the peak position ratio $f_{\text{max}}/f_{\alpha}$. Green cross is the glycerol ratio\cite{gainaru2019spectral} and the blue dashed line depicts the ratios of Debye- to $\alpha$-process peaks from BDS measurements of monohydroxy alcohols.\cite{bierwirth2018communication}} 
  \label{fig_envelope}
\end{figure}
Even more recently, Gainaru\cite{gainaru2019spectral} showed that the above correlation can be represented by a scaling relation, where the dielectric spectra of different polar molecules scale onto a universal "envelope" curve. Evaluating each point of this envelope with respect to its slope, from which $\beta_{{K}}$ is deduced, and its amplitude, he could reproduce the correlation of $\Delta \epsilon$ and $\beta_{K}$ found by Paluch {\it et al.}. That means that, in absence of a secondary relaxation, all molecular glass formers should share a common high frequency dielectric slope and just the main relaxation peak is located at different points on the envelope depending on the polarity of the molecule. In the light of the presented findings it is interesting to figure out, if this envelope is consistent with the idea of an additional Debye process being present due to interacting dipoles. Thus, we test this picture for TBP and its mixtures with n-pentane in Fig.~\ref{fig_envelope}.

In this figure the envelope is taken from Ref.~\citenum{gainaru2019spectral} (orange dashed line) and the spectra of neat and mixed TBP are scaled horizontally and vertically in such way that a good overlap with the envelope is achieved. Due to the pronounced secondary relaxation, the overlap could be obtained in a small frequency window only, but since the slope of the envelope changes smoothly over the whole frequency range until it reaches $\omega^{-1}$ at very low frequencies, the superposition is unambiguous.      
Solid black lines are fits, which are obtained for the BDS data -- after superposition on the envelope -- by adding a Debye process to the DDLS fit. The red dashed line is obtained by adding an arbitrary Debye process to the DDLS with the restriction on $\Delta \epsilon$ and $\tau_D$ such that the resulting spectra resembles the envelope. This procedure shows on the one hand that the data for neat and diluted TBP is in accordance with the Gainaru envelope and thus with the Paluch correlation, and on the other hand that the proposed picture of $\alpha$- plus additional Debye process can reproduce the envelope for TBP, its mixtures and also a hypothetical, even  stronger Debye processes. 

The non-trivial point about this representation is, that only certain combinations of $\Delta \epsilon_{\text{D}}$ and $\tau_{\text{D}}$ with respect to a given $\Delta \epsilon_\alpha$ and $\tau_\alpha$ can reproduce the envelope. The dashed black line in Fig.~\ref{fig_envelope} is proportional to $f^{-1}$ and placed in such way that it passes through all the peak maxima. 
In accordance, the inset shows the peak height from BDS divided by the one from DDLS as a function of the peak position from BDS divided by the one from DDLS.
 The black dashed line represents a power law $\propto (f_{\text{max}}/f_{\alpha})$, which fits the data of pure and diluted TBP. The green cross is obtained from BDS and DDLS data of glycerol\cite{gainaru2019spectral} in the same way as for TBP. It coincides perfectly with the black dashed line, indicating the same behavior for glycerol, a hydrogen bonding liquid, and TBP. 
The correlation found for the Debye- and $\alpha$-process of a homologous series of monohydroxy alcohols (MAs) and their mixtures\cite{bierwirth2018communication} is also shown as a blue dashed line with the same slope but different intercept. Thus, at the same peak height, the Debye process is more separated from the $\alpha$-process in MAs than in other liquids, but $\frac{\epsilon''_{\text{max}}}{\epsilon''_{\alpha}} \propto \frac{f_{\alpha}}{f_{\text{max}}}$ holds in both cases. 

The fact that the separation of the Debye- and $\alpha$-process increases with increasing $\Delta \epsilon$ can be understood  qualitatively by combining the static and dynamic part of the D\'ejardin theory.\cite{dejardin2018calculation, dejardin2019linear} Both parts incorporate the interaction parameter $\lambda \propto \frac{\rho_0\mu^2}{k_B T}$, where $\rho_0$ is the number density, $\mu$ the dipole moment and T the temperature. In the static part of the theory, the Kirkwood correlation factor $g_K>1$ increases with increasing $\lambda$, leading to a higher $\Delta \epsilon$ through the Kirkwood-Fröhlich equation.\cite{boettcher1978theory1}
In the dynamic part, the peak appearing for $g_K>1$ shifts to lower frequencies for increasing $\lambda$, which renders the combined peak of $\alpha$- and Debye process narrower, as seen in figure \ref{fig_envelope}. Together, the experimental observed correlation between the dielectric strength $\Delta \epsilon$ and the dielectric loss peak shape parameter $\beta_{\text{K}}$ is qualitatively rationalized by this theory. 

In conclusion, we have shown that in the dielectric spectrum of the van der Waals liquid TBP, in addition to the $\alpha$-process probed by depolarized dynamic light scattering, a Debye process is present, which arises due to dipole-dipole interactions, in line with a recent theory by D\'ejardin \emph{et al.}. By dilution with the apolar solvent n-pentane, the dipole-dipole interactions could be reduced and the Debye process vanished. We showed that phenomenologically this picture is able to rationalize correlations between the dielectric strength and the shape parameter $\beta_K$, which were found recently for a large number of glass forming liquids.\newline 

We are indebted to Pierre-Michel D\'ejardin, University of Perpignan, and Yann Cornaton, University of Strasbourg, for fruitful discussions. We are also grateful to Ernst Rössler, Bayreuth, for making the dielectric time domain setup available to us.
Financial support by the Deutsche Forschungsgemeinschaft under grant No. BL~1192/3 is gratefully acknowledged.

\bibliography{bib_fp.bib}

\begin{thebibliography}{41}
\bibitem [{Tg_()}]{Tg_idealMix}%
  \BibitemOpen
  \href@noop {} {}\bibinfo {note} {The glass transition temperature $T_g$ of
  n-Pentane is approximated by extrapolation the $T_g$s of the mixtures (taken
  as the temperature at which $\tau$=100s) to neat n-Pentane by assuming ideal
  mixing.}\BibitemShut {Stop}%
\end{thebibliography}

\end{document}